# *Effective suppression of dark counts in superconducting microstructures with grid of pinholes in a magnetic field*

Dong Zhu, Ilya Charaev, Andreas Schilling

Department of Physics, University of Zurich, Winterthurerstrasse 190, CH-8057 Zurich, Switzerland

*In a magnetic field, vortices significantly contribute to the dark counts of single-photon detectors made of superconducting wires, and they are also limiting the critical current of such devices. To address this issue, we prepared superconducting microwires with a pinhole grid from WSi thin films and report on corresponding critical-current and count-rate measurements in an external magnetic field B. When compared to corresponding devices without pinholes, the critical current only weakly depends on the magnetic field at B < 16 mT and it is even larger already at B > 10 mT. Moreover, dark counts are not only suppressed in zero field, but particularly in magnetic fields B < 16 mT, while photon counts are virtually field insensitive in the same range of the magnetic field.*

## Introduction

After more than two decades of research since the first experimental report in 2001[1], superconducting nanowire single-photon detectors (SNSPDs) have made great progress in many aspects of performance, such as high detection efficiency[2-6], low dark counts[3, 7, 8], high temporal resolution[9], kilo-pixel array[10] and a wide range of useful wavelengths[5, 11-13]. Recently, research in related photon detectors has entered a further phase as several research groups have reported the detection of single photons in superconducting wide microwire detectors made of various of different materials, such as NbN[14], WSi[15] and MoSi[16]. Contrary to the well-known hot spot-based theoretical model, which considers the width of superconducting wires only up to a few hundred nanometers, these experimental advances suggest completely different detection mechanisms, e.g., a vortex-assisted scenario in which vortex and anti-vortex pairs are generated whenever a photon is absorbed in the microwire, and their motion leads to a switching to the normal state along with the generation of a measurable voltage pulse [14, 17]. An increase of the wire width $w$ to the micron scale offers more freedom to design new detectors based on current nano- and even microfabrication techniques.

It is well known that defects, such as artificial holes, can trap vortices and hinder their motion, thereby increasing the current-carrying capability of a superconducting film[18-22]. This phenomenon can be used for changing the properties of superconducting microwires and detectors[23]. The effect of a single hole on the critical current of a NbN-based microbridge was recently investigated, and corresponding experiments showed that the application of an external magnetic field $B$ has only a weak effect on the critical current $I_c$, leading to a plateau in $I_c(B)$. This has been taken as an evidence for vortex and anti-vortex nucleation near the hole[24].

Such a plateau in $I_c(B)$ modifies the properties of a superconducting microwire single-photon detector (SMSPDs) by making it much less sensitive to the presence of external magnetic fields in comparison to ordinary SMSPDs, where $I_c$ decreases rapidly when a magnetic field is applied[25]. As we demonstrate below, an SMSPD made of a 2-nm thin WSi film with a pinhole grid not only shows a plateau but also an enhanced $I_c$ in a magnetic field, leading to stable behaviour with respect to critical current, dark counts and photon counts in a certain range of the magnetic field.

**Experimental**

The fabrication of our devices started with the preparation of a 2-nm thin WSi film by co-sputtering from W and Si targets in argon atmosphere a 3-inch silicon wafer with a 50 nm thermally oxidized $SiO_2$ layer. The stoichiometry of the WSi thin film was determined by EDX to $W_{0.59}Si_{0.41}$ with a sheet resistance of ≈ 1000 Ω/□. The 1-µm wide meander-shaped wires were patterned by electron-beam lithography at 20kV acceleration voltage on 60 nm PMMA A2 resist, followed by reactive ion etching in $SF_6$. Figure 1 shows two types of structures, one with and one without pinholes. Each of the 90 µm long wires was 1 µm wide, and 61 wires were arranged into a meander structure with a filling factor of 50% and an active area of 90x120 µm². Figure 1(a) shows the meander structure without pinholes and Figure 1(b) the corresponding structure with pinholes. In the straight part we designed a triangular hole arrangement with 200 nm between the center of the holes, while we have chosen a hexagonal geometry in the meander turns. The nominal hole diameter was 100 nm by design. However, the effective hole diameters of the actual device scattered by up to 30% around the nominal value due to the structuring process (e.g., dose errors and instability of the electron beam lithography technique). As we shall show below, however, the "effective" width of the lines with holes as deduced from the normal-state resistance is in agreement with the design value of the width, which is reduced from 1 µm by ≈ 3x100nm = 300 nm to ≈ 0.7 µm. The measurements were performed in a dilution refrigerator with a base temperature of approximately 200 mK. The bias current and voltage pulses were applied and amplified by a DC source and amplifier, respectively. Optical experiments were carried out by guiding 850 nm and 1310 nm wavelength light from a laser source through a standard SMF 28e optical fiber leading to the cold stage. We attenuated the incident light by four orders of magnitude of the total laser power to reach the single-photon regime.

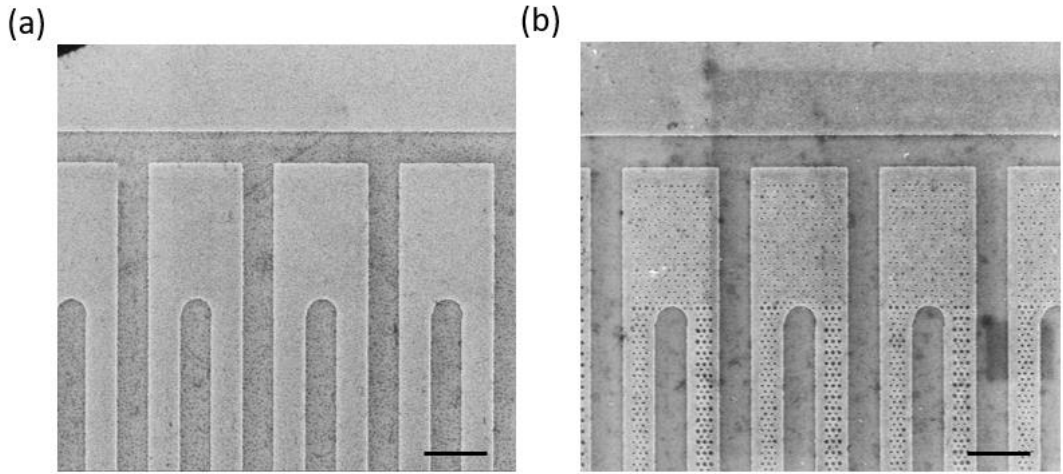

*Fig. 1 SEM sample images. (a) 1-µm wide meander with a filling factor of 0.5, the scale bar is 2 µm; (b) 1-µm wide meander with hole design, the effective diameter of the holes varies within up to 30% around the nominal diameter of 100 nm, the scale bar is 2 µm.*

**Results**

The temperature dependence of the resistances of both samples are shown in Figure 2. The presence of pinholes reduces the meander cross section and therefore increases the normal-state resistance, and we obtain a ratio

$R_{\mathrm{mea}}/R_{\mathrm{pin}} \sim 0.7$, where $R_{\mathrm{pin}}$ and $R_{\mathrm{mea}}$ denote the normal state resistances at room temperature for the pinhole meander and the normal meander, respectively. Due to processing-induced degradation[19], the superconducting critical temperature shows a slight difference, i.e., a $T_c \sim 2.3$ K for the pinhole meander and a slightly higher $T_c \sim 2.4$ K for normal meander, as shown in the inset of Figure 2.

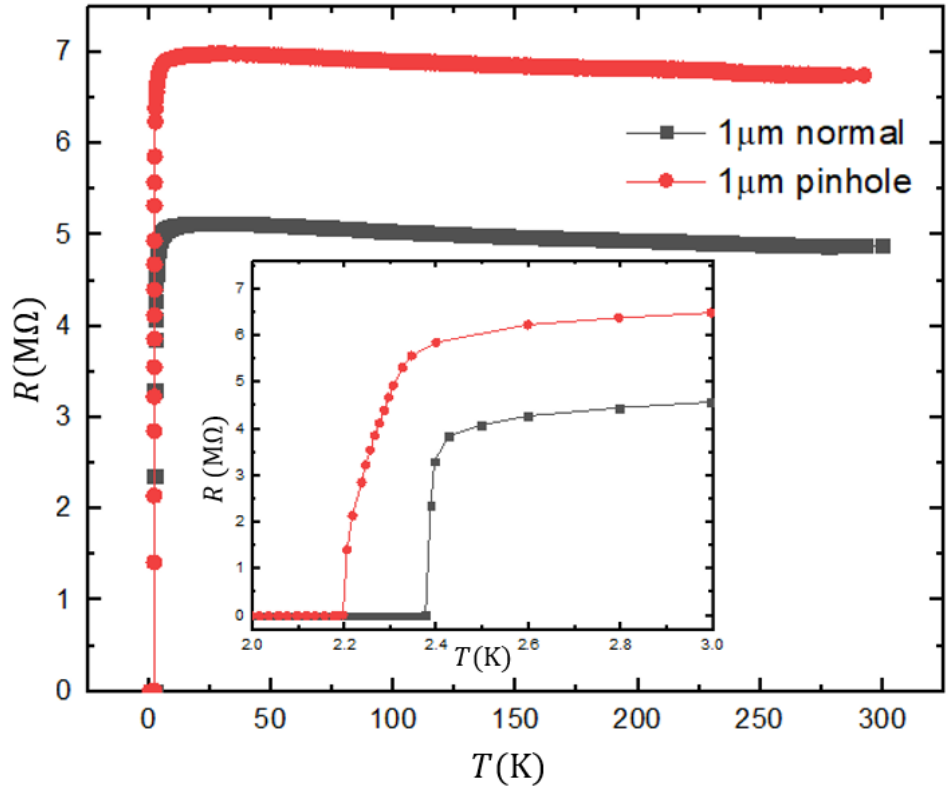

*Fig. 2  The temperature dependence of the resistances of the meanders with("pinhole") or without ("normal") pinholes. The inset shows the same data near the superconducting transition.*

The difference in the critical-current characteristics between the two detectors in magnetic fields between ≈ -40 mT and +40 mT is depicted in Figure 3. The black squares, taken from corresponding data of the normal meander, reproduce the suppression of the critical current in a magnetic field which is caused by the Meissner current that lowers the barrier threshold for vortex penetration at the meander edges, and is limiting the current-carrying capability of the device[25]. The normal meander has a maximum of $I_{c,max} \sim 15.2$ µA, which can be rescaled by the factor 0.7 to $I^*_{c,max} \sim 10.6$ µA (blue triangles) of a device with equal current density but an effective wire width that is comparable to that of the pinhole meander. The critical current decreases rapidly with increasing magnetic field, with a half width at half maximum value $\Delta B_{HWHM} \sim 10$ mT. In contrast, the critical current of the pinhole meander has a very different magnetic-field dependence (red circles in Fig. 3), with a reduced maximum of $I_{c,max} \sim 8$ µA on the one hand, but a virtually unchanged value in a magnetic $B \approx \pm 19$ mT around $B = 0$ on the other hand. Moreover, $I_c(B)$ strongly exceeds the corresponding values measured in the normal meander structure already at $B > 10$ mT.

We have also compared the dark-count rates of the two detectors by varying the bias current and the magnetic field. The Figures 4 (a, b) show the dark-count rate (DCR) vs. bias current $I_B$ for normal and pinhole meanders at different magnetic fields, respectively. For the normal meander, the bias current was rescaled according to the resistance ratio in the normal state to have comparable effective wire widths and current densities. In the normal meander, there is a strong shift in the corresponding DCR data towards low bias currents with increasing magnetic field (Fig. 4a). By contrast, the corresponding DCR dependences of the pinhole meander show a comparably small shift, particularly for B < 16 mT (Fig. 4b) while the shift also increases in larger fields up to 30 mT, which is the largest magnetic field investigated here. This behaviour can also be seen clearly when measuring the DCR vs. magnetic field by fixing the bias currents and varying the magnetic field around zero (Figs. 4c and 4d).

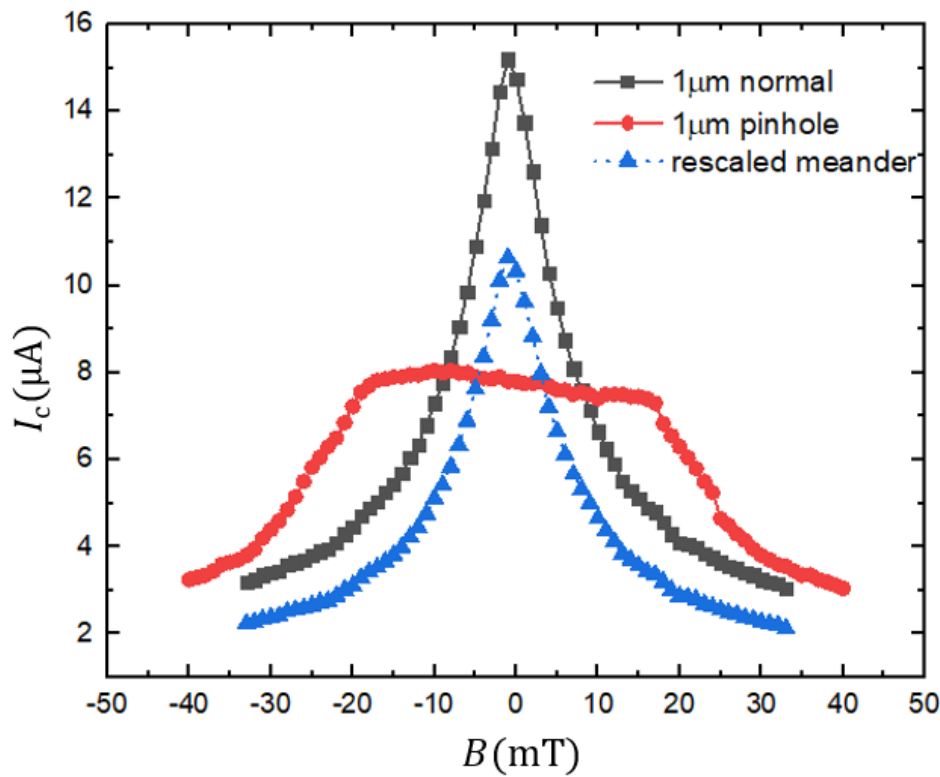

*Fig. 3 Magnetic-field dependence of the critical current $I_c$. Red, black and blue symbols indicate pinhole meander, normal meander and corresponding rescaled data, respectively, where the rescaling accounts for the correction due to the different effective wire widths (see text).*

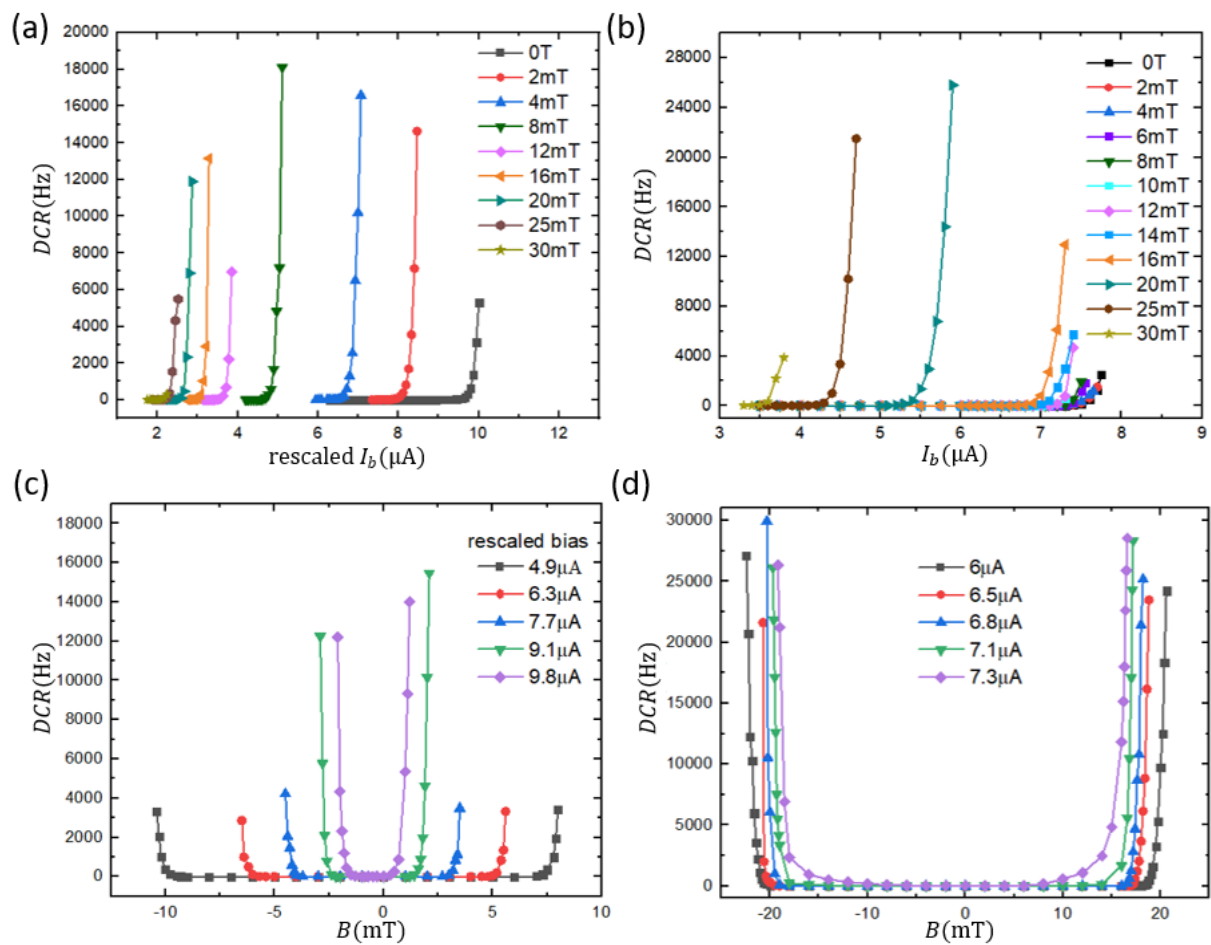

*Fig. 4 Dark-count rates as functions of bias current and magnetic field. (a) DCR vs. rescaled bias current for the normal meander for fixed magnetic fields; (b) corresponding DCR vs. bias-current data for the pinhole meander; (c) DCR vs. magnetic field for the normal meander for fixed rescaled bias currents; (d) DCR vs. magnetic field for the pinhole meander for selected constant bias currents.*

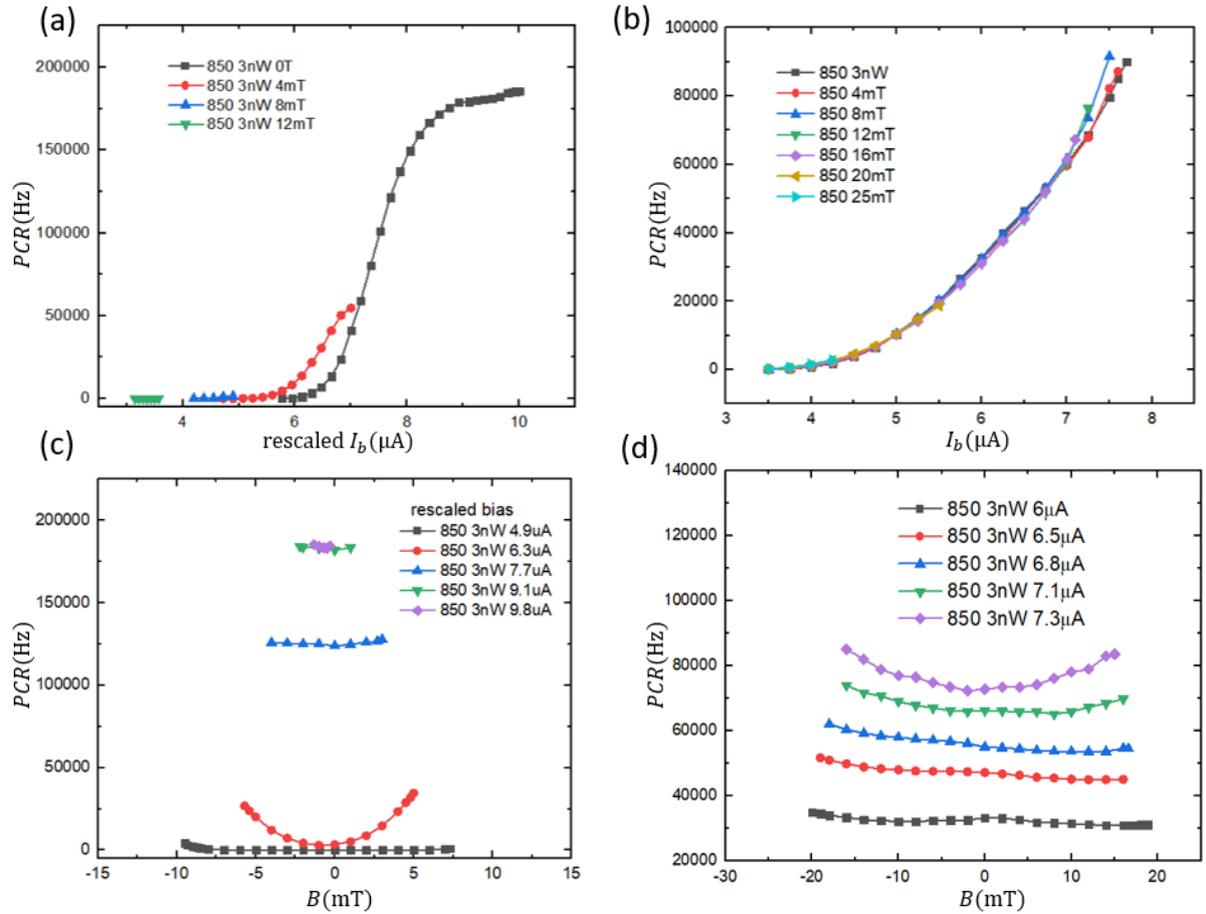

*Fig. 5 Photon count rates (PCR) vs. bias current and magnetic field under light illumination with 850 nm wavelength. (a) PCR of the normal meander vs. rescaled bias current for fixed magnetic field; (b) PCR vs. bias current of the pinhole meander; (c) PCR of the normal meander vs. magnetic field for several rescaled bias currents; (d) corresponding PCR data of the pinhole meander.*

As expected from the data displayed in Fig. 3, the magnetic-field window where the DCR is negligible in the normal meander is narrowing with the bias current increasing towards the zero-field critical current (Fig. 4c). The corresponding magnetic-field window for the pinhole meander is much wider upon approaching the zero-field critical current plateau (Fig. 4d), thereby reflecting again the beneficial influence of the pinholes on the dark-count rate.

We have measured the photon count rates (PCR) of both detectors, as shown in Figure 5. These data have been corrected by the expected dark counts as derived from our current and field dependent DCR measurements. Figure 5(a) shows the PCR of the normal meander at an illumination with light of 850 nm wavelength and a total laser power of 3 nW, with a tendency to form a clear plateau at bias currents near 9 µA in zero magnetic field, indicating that the internal detection efficiency becomes saturated. The detection performance decreases rapidly with the application of a magnetic field[26], and full saturation for $I_b < I_c$ could not be achieved. Scanning the magnetic field for fixed rescaled bias currents as shown in Fig. 5(c) reveals that the PCR at a large bias current close to the zero-field critical current is only weakly field dependent, although the useful magnetic-field window is strongly narrowing as we have shown in Fig. 3. The PCR of the pinhole meander, by contrast, does not show any plateau with varying bias current (see Fig. 5(b)), meaning that the internal efficiency is not saturated before the meander wire switches to the normal state. However, it exhibits virtually no performance degradation in magnetic fields below 16 mT (Fig. 5(b)), which also becomes obvious from corresponding measurements at constant current and varying magnetic field as shown in Fig. 5(d). We note that the slight asymmetry upon changing the direction of the magnetic field in Figs. 4 and 5 must be caused by a not quite perfectly symmetrical geometry of the meanders.

Finally, we examined the single-photon detection behaviour of the pinhole meander at a lower photon energy. In Fig. 6(a) we show the total count rate (including dark counts) upon illumination with light of 1310 nm wavelength and a total laser power of 1 nW. Dark counts manifest themselves in a sudden upturn of the total count rate as a function of bias current, as can be clearly seen in the data measured in 20 mT and 24 mT. In agreement with the data shown in 5(b), the count rates virtually do not change for bias currents below ≈ 7 µA and magnetic fields below 16 mT. We have therefore chosen 6.8 µA and 7.3 µA as two bias currents to test single photon response (see Figs. 6 (b,c))[15]. The linearity of the count-rate as a function of laser power suggests that the pinhole meander is working in the single-photon regime, at least for the magnetic fields at and below 12 mT.

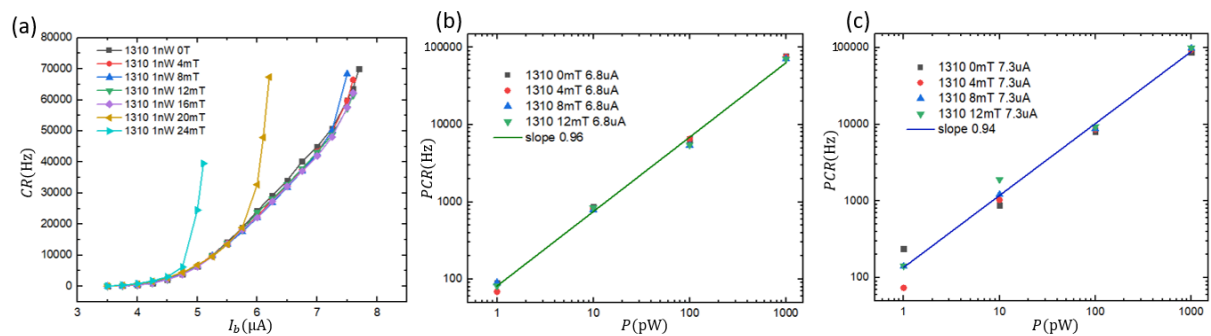

*Fig.6 Single-photon detection behaviour of the pinhole meander detector at 1310 nm wavelength. (a) Count rate vs. bias currents upon changing the magnetic field. (b) Photon count rate vs. total laser power for different magnetic fields while fixing the bias current at 6.8 µA and (c) at 7.3 µA.*

**Conclusions**

To summarize, we investigated the properties of a superconducting microwire pinhole meander, showing single-photon detection and dark-count properties that are significantly different from those of normal microwire meander detectors[14-16, 23]. This pinhole meander exhibits a robust critical current against magnetic field within a certain magnetic-field window spanning ≈ +/- 16 mT [24, 27]. This has been attributed in single superconducting microwires to vortex-anti-vortex generation near the pinholes rather than at the wire edges for sufficiently small magnetic fields and resulting small Meissner screening currents [23]. Remarkably, dark counts are also suppressed, suggesting that the holes along the wire are limiting vortex motion and are thereby reducing the tendency to spontaneously form a normal conducting region. Photon-count rates are virtually field independent in this range of magnetic field, and the detector has been shown to work in the regime of single-photon detection. The fact that vortex motion is impaired is supported by the enhancement of the critical current for $B$ > 10 mT by the presence of the artificial pinholes in the otherwise weakly-pinning WSi material.

Although we did not observe saturated photon counting behaviour, our results suggest that geometric modifications of microwire single-photon detectors can dramatically alter the physical properties and potentially improve photon detection performance, particularly in a magnetic environment. We mention that the chosen pinhole density corresponds to a matching field (i.e., the magnetic field where the vortex density matches the hole density) of ≈ 45 mT. Although this is in the region of the magnetic fields applied in this study, it has been suggested that it may be the hole size that determines the width of the magnetic field window with a plateau in the critical current of superconducting microwires [23]. In this sense, there is still much room for studying the effects of size, density, and the arrangement geometry of pinholes to further improve the detector performance of such superconducting microwire detectors.